\documentclass[fleqn,10pt]{wlscirep}
\usepackage[utf8]{inputenc}
\usepackage[T1]{fontenc}
\title{{Multi-Plane Vision Transformer for Hemorrhage Classification Using Axial and Sagittal MRI Data}}

\author[1,2*]{Badhan Kumar Das}
\author[3]{Gengyan Zhao}
\author[3]{Boris Mailhe}
\author[3]{Thomas J. Re}
\author[3]{Dorin Comaniciu}
\author[3]{Eli Gibson}
\author[2]{Andreas Maier}

\affil[1]{Digital Technology and Innovation, Siemens Healthineers, Erlangen, Germany}
\affil[2]{Pattern Recognition Lab, Department of Computer Science, Friedrich-Alexander-Universität Erlangen-Nürnberg, Erlangen, Germany. }
\affil[3]{Digital Technology and Innovation, Siemens Healthineers, Princeton, New Jersey, United States}

\affil[*]{badhankumar.das@siemens-healthineers.com}


\begin{abstract}

Identifying brain hemorrhages from magnetic resonance imaging (MRI) is a critical task for healthcare professionals. The diverse nature of MRI acquisitions with varying contrasts and orientation introduce complexity in identifying hemorrhage using neural networks. For acquisitions with varying orientations, traditional methods often involve resampling images to a fixed plane, which can lead to information loss. To address this, we propose a 3D multi-plane vision transformer (MP-ViT) for hemorrhage classification with varying orientation data. It employs two separate transformer encoders for axial and sagittal contrasts, using cross-attention to integrate information across orientations. MP-ViT also includes a modality indication vector to provide missing contrast information to the model. The effectiveness of the proposed model is demonstrated with extensive experiments on real world clinical dataset consists of 10,084 training, 1,289 validation and 1,496 test subjects. MP-ViT achieved substantial improvement in area under the curve (AUC), outperforming the vision transformer (ViT) by 5.5\% and CNN-based architectures by 1.8\%. These results highlight the potential of MP-ViT in improving performance for hemorrhage detection when different orientation contrasts are needed.

\end{abstract}
\begin{document}

\flushbottom
\maketitle
%
%
\thispagestyle{empty}


\section*{Introduction}

Intracranial hemorrhage (ICH) is a critical medical condition which, on suspicion, requires rapid and accurate detection and characterization.  The long-time reference standard technique for its detection is computed tomography (CT).  Magnetic resonance imaging (MRI) has been shown to be at least as sensitive to CT in detecting ICH while offering more detailed characterization of this finding, particularly of its chronicity (acute vs subacute vs chronic nature), while avoiding patient exposure to ionizing radiation \cite{romanova2014magnetic}.  Furthermore, MRI offers superior detection of other critical intracranial findings, in particular, ischemia \cite{von2002mri}.  Still, MRI has not replaced CT for most imaging of patients with suspected ICH. This is due, in part to issues of availability and longer scan times of MRI over CT, an issue being actively addressed through new technologies.  A further issue impeding MRI's adaptation in ICH diagnosis is the higher complexity involved with interpreting MRI results over those of CT.  The application of machine learning and deep learning methods can be used to significantly reduce the complexity of such interpretation while enhancing accuracy and efficiency in this context.

In the rapidly evolving field of deep learning and computer vision, the application of deep learning techniques for medical image processing have made substantial strides, revolutionizing tasks such as image classification \cite{ismael2020medical,rahmat2018chest}, object detection \cite{jaeger2020retina,li2019clu}, and segmentation \cite{malhotra2022retracted,roth2018deep}. Recently, the successes of vision transformers \cite{dosovitskiy2020image} on natural images, have encouraged researchers to query further the application of self-attention in medical imaging in order to capture long-range dependencies between pixels. Additionally, transformers have achieved comparable performance to state-of-the-art convolutional neural networks (CNNs) on medical image classification \cite{manzari2023medvit,dai2021transmed}, detection \cite{he2023transformers,shou2022object}, segmentation \cite{hatamizadeh2021swin,hatamizadeh2022unetr,zhou2023nnformer} and reconstruction \cite{feng2022multi,luthra2021eformer}.

In clinical practice, MRI scans are often performed with relatively thick slices to speed up the imaging, leading to anisotropic images with fewer slices that may not be sufficient to detect abnormalities. To enhance the diagnostic accuracy, clinicians frequently use multiple planes (axial, sagittal, and coronal) to get a comprehensive view of anatomical structures. Different planes reveal various aspects of anatomical structures, and some abnormalities or hemorrhages might be more visible in a particular plane due to their orientation. However, in both CNNs and vision transformer models, the input images are needed to be processed to a specific plane due to architectural constraints. Consistent alignment allows the neural network to effectively learn spatial relationships and integrate information from different channels, which is essential for accurate analysis of volumetric data. Despite the widespread adoption of using CNN and ViT for medical image classification tasks, concerns linger about potential information loss or distortion, especially when resampling MRI images from one orientation to another. For anisotropic images this change in orientation may need extensive down sampling/up sampling which can lead to significant information loss. 

Deep learning models have proven effective for classifying hemorrhages, such as in CT brain hemorrhage\cite{jnawali2018deep} and intestinal hemorrhage\cite{pannu2020deep}. Recently, transformer-based architecture has been used by Adel et al.\cite{elzemity2023transformer} for intracranial hemorrhage classification and detection. Sharma et al\cite{sharma2023novel}, used vision transformer with residual in self-attention for brain tumor classification using MR images. ScopeFormer\cite{barhoumi2021scopeformer} used a hybrid CNN-Transformer architecture for intracranial hemorrhage detection.  In another study, multi-stage transformer fusion is used for efficient hemorrhage classification\cite{wang2024multi}. However, all these studies do not address how to handle differences in image orientation. Existing methods require resizing the anisotropic input images from different orientations to the same resolution, which may introduce unnecessary resolution reduction and inaccurate interpolation. Another issue with real-world clinical datasets is that they often have inconsistent contrast across cases. 

This motivates the exploration of innovative methodologies to alleviate the necessity for changing orientations and optimizing the performance and robustness of deep learning models. Building on the successes of vision transformers in medical imaging, we extend this paradigm to propose a novel approach to hemorrhage classification using the multi-plane vision transformer (MP-ViT) architecture with axial and sagittal brain MRI data. Traditionally, medical imaging analyses have focused on individual planes, limiting a comprehensive understanding. Our approach consolidates information from multiple imaging perspectives, providing a more comprehensive representation of hemorrhagic regions and enhancing the model’s discriminative power. We also introduce a modality indication vector inspired by dynamic filter ideas from ModDrop++ \cite{liu2022moddrop++} to address the challenge of missing contrasts in the input data. This vector serves as a guide for the Transformer model, enabling it to better handle variations in imaging modalities and effectively compensate for the absence of certain contrasts. By incorporating this modality indication vector, our approach ensures that the model maintains its performance and accuracy even when faced with incomplete or heterogeneous input data, ultimately leading to more robust and reliable hemorrhage classification results.

\section*{Methods}

\subsection*{Dataset and Pre-processing}

This retrospective study is compliant with the Health Insurance Portability and Accountability Act (HIPAA). The paper describes scientific research using retrospectively acquired anonymized data. The anonymization was performed in accordance with applicable laws and regulations before secure transfer was made to Siemens Healthineers for the study. Use of the data followed all the applicable license terms. The study does not involve any clinical or human subject research component to it. As such, IRB approval and informed consent are not applicable because there are no data privacy issues and no patients were impacted by the research (did not affect treatment or diagnosis). The research was conducted conforming to the appropriate scientific practices and in accordance with the relevant guidelines and regulations of the institution conducting the experiments.


Our dataset consists of 10,084 training, 1289 validation, and 1496 test subjects of 3D MRI data. The dataset was acquired from scanners produced by three manufacturers (GE, Philips, and Siemens Healthineers). Each of the subjects was labeled based on re-analysis of each radiology report by one of three board-certified radiologists (A.D., K.N. and T.J.R with at least 10 years of experience each). Subarachnoid, intraparenchymal, and extra-axial hemorrhages were labelled as ICH. Any descriptions related to chronic hemorrhages were excluded from this. Both acute and subacute intracranial hemorrhages are considered as positive classes for this task. The positive-to-negative sample ratio for the entire dataset approximates 1:13. For this experiment, we followed an inclusion criterion that required the presence of axial T2 FLAIR, axial apparent diffusion coefficient (ADC), and axial Trace-weighted imaging contrasts. In addition, we considered four other contrasts as optional: axial T2-weighted, axial gradient echo (GRE) (T2*-weighted), sagittal T1-weighted, and axial susceptibility weighted imaging (SWI) (T2*-weighted). The MRI imaging parameters are provided in the supplementary materials.
 
 As a part of data pre-processing axial images are resized to a fixed dimension [256 × 256 in-plane and 32 out-of-plane] and sagittal images are resized to [256 × 32 × 256]. For conventional methods which cannot process multi-plane data, we change the orientation of the sagittal images to axial plane. A linear transformation is used to normalize the image intensities of each contrast between 0 and 1.

\subsection*{Network Architecture}

\begin{figure}[ht]
\includegraphics[width=\textwidth]{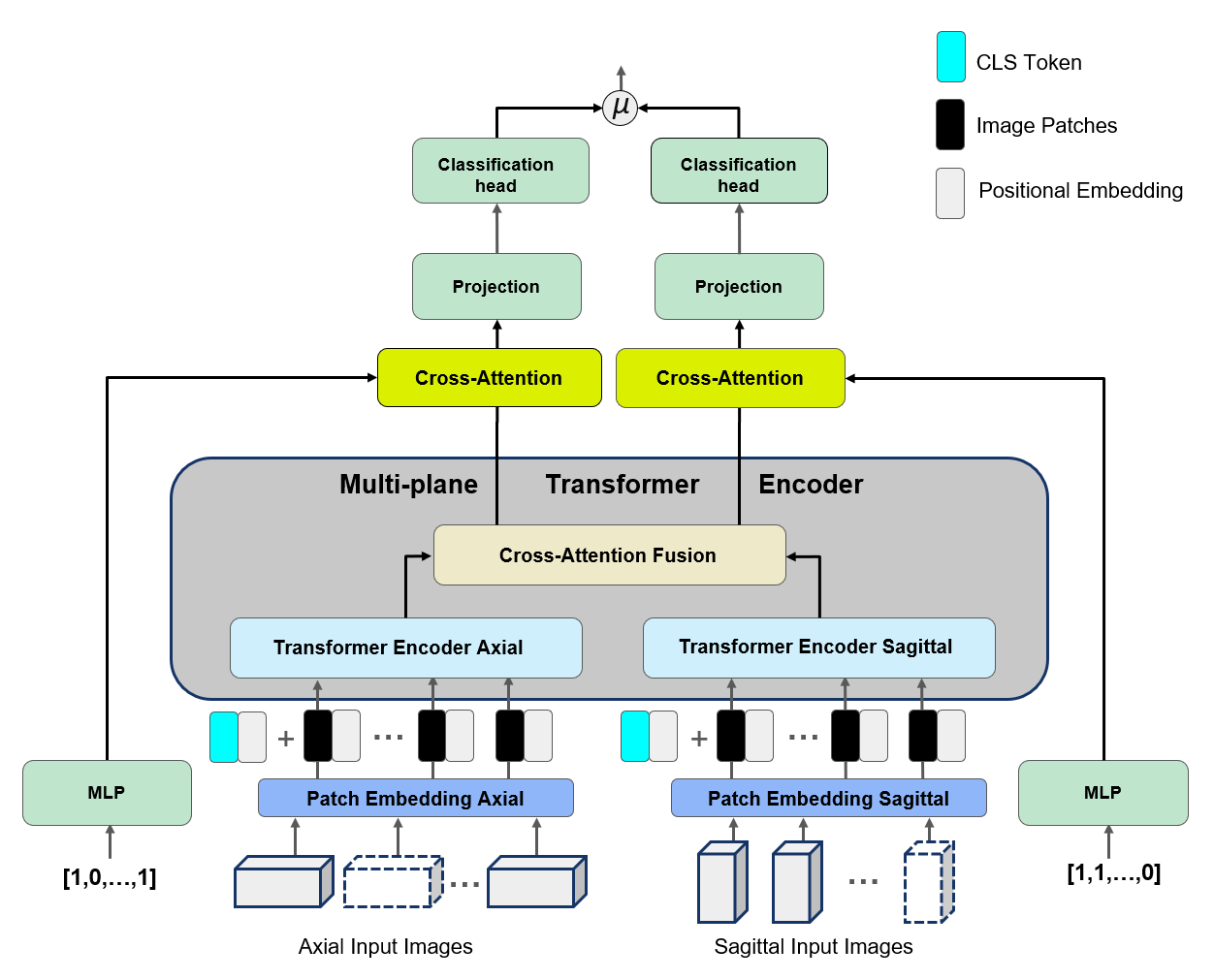}
\caption{Overview of multi-plane vision transformer with axial and sagittal transformer encoder} \label{architecture}
\end{figure}

The proposed architecture is built on top of vision transformer \cite{dosovitskiy2020image}, self-attention \cite{vaswani2017attention} and cross-attention \cite{chen2021crossvit} mechanism.The detailed design of the multi-plane vision transformer is illustrated in Figure \ref{architecture}. This architecture is designed in such a way that it can handle variable numbers of axial and sagittal images as input. The whole design consists of four modules:

\begin{itemize}
\item The first module consists of two different transformer encoders for axial and sagittal planes. Non-overlapping patches are created from axial and sagittal 3D images separately. Then these tokens are fed into corresponding axial and sagittal transformer encoders. 
\item The second module has a cross-attention fusion mechanism to relate the information from each branch to another. 

\item The third module has cross-attention between modality indication vector and transformer output to provide information about used modalities.

\item Finally, two different classification heads are used to identify hemorrhage. The mean of these two heads is used as the final output of the model.

\end{itemize}

The detailed design of each component is introduced below.

\subsection*{Axial and Sagittal Transformer Encoder}

Multi-plane vision transformer is a flexible transformer architecture which comprises of two branches: one designated for processing axial input images and the other for handling sagittal images. Within each branch, images undergo resampling to conform to the corresponding anisotropic image grid, such as (256, 256, 32) for axial images and (256, 32, 256) for sagittal images. Subsequently, the resampled images within each branch are concatenated along the channel dimension, divided into patches, and tokenized by the corresponding patch embedding layer. 

Let $I \in \mathbb{R}^{H \times W \times D \times C}$ represent the input image, where $H$, $W$, and $D$ are the height, width, and depth dimensions, respectively, and $C$ is the number of channels. Subsequently, the new image $I$ is divided into non-overlapping patches denoted in equation \ref{eq:patch}. where $P$ signifies the resolution of each patch ($16 \times 16 \times 16$ in our case), and $N$ represents the length of the sequence. This can be expressed mathematically as shown in equation \ref{eq:N}. Each patch is then flattened and projected into a transformer-dimensional token by a patch embedding layer as shown in equation \ref{eq:token}.

\begin{equation}
\text{Patches} \in \mathbb{R}^{N \times P^3}
\label{eq:patch}
\end{equation}

\begin{equation}
N = \frac{H \times W \times D}{P^3}
\label{eq:N}
\end{equation}

\begin{equation}
\text{Token}_{ijk} = \text{Embedding}(I_{i,j,k})
\label{eq:token}
\end{equation}

where $I_{i,j,k}$ is the $(i,j,k)$-th patch in the input image.

Similar to the original BERT \cite{devlin2018bert} and vision transformer \cite{dosovitskiy2020image} (ViT) models, an additional classification (CLS) token is added to the sequence of tokenized patches. This token is used for the final classification task after the transformer layers have processed all the tokens. Positional embeddings are then added to each token, including the CLS token  as the Transformer is position-agnostic. This step ensures the model retains information about the spatial relationships between patches.

\begin{equation}
\text{Token}_{all} = \text{Token}_{CLS} + \text{Token}_{ijk}
\end{equation}

\begin{equation}
\text{Token}_{all}^{\text{pos}} = \text{Token}_{all} + \text{PositionalEmbedding}_{all}
\end{equation}

This process precedes feeding the tokens into the respective branch's transformer. Notably, this approach treats axial and sagittal images as two distinct modalities, obviating the need to resample sagittal images based on the grid of axial images or vice versa.

\subsection*{Cross Attention Fusion}



\begin{figure}[ht!]
\centering
\includegraphics[width=1.0\textwidth]{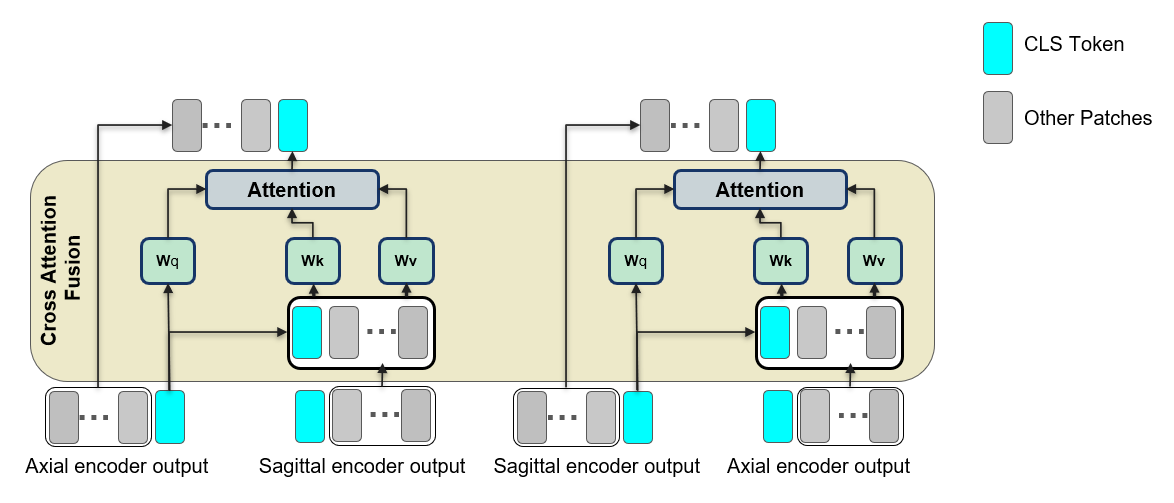}
\caption{Overview of the cross attention fusion used in MP-ViT.  In this block, the CLS token of the axial encoder input performs attention with the other tokens of the sagittal encoder. Similary, this process is done also with the CLS token of sagittal encoder input with the other tokens of the axial encoder. Here \textbf{W}q, \textbf{W}k and \textbf{W}v are learnable matrices to create query, key and values for attention.} \label{crossfusion}
\end{figure}

Post-transformer processing involves establishing inter-branch relationships through cross-attention fusion, which facilitates the extraction and integration of pertinent information from each branch. In this cross-attention fusion, CLS token of one branch is used with other tokens of the other branch as shown in Figure \ref{crossfusion} in order to fuse each transformer encoder's features more efficiently and effectively than all-attention fusion \cite{chen2021crossvit}. CLS token is used to exchange information among the patch tokens from  the other branch and then project it to its own branch. As the CLS token already learns abstract information from it's own branch, combining it with patches from other branch helps to learn information from different planes.

\subsection*{Cross-Attention with Modality Vector}


To address the challenge of missing any contrast within each branch of the proposed model we utilize a modality indication vector. This vector is constructed as a binary vector, with its length equal to the number of all possible modalities for the respective branch. The presence or absence of each modality is represented by the values 1 and 0, respectively. Instead of a dynamic head similar to ModDrop++\cite{liu2022moddrop++} a multilayer perceptron is used to embed the modality indicator vector into a sequence of query vectors with the dimension of the transformer. This approach simplifies the integration of modality information, reducing the computational complexity associated with dynamic head generation. These query vectors are then employed to conduct cross-attention with the transformer's output for each branch. The resulting cross-attention output serves as the input for the subsequent stages of the network. This approach ensures that information pertaining to missing modalities is provided to the model, thereby enhancing the model's robustness in handling incomplete multi-modality input data. 

Given a query sequence $\mathbf{q}$ created from the projection of modality indication vector and key-value sequence $\mathbf{k}$ and $\mathbf{v}$, the cross-attention mechanism can be defined as follows:

\[
\mathbf{q} = \text{MLP}(y) \mathbf{W}_Q, \quad \mathbf{k} = x \mathbf{W}_K, \quad \mathbf{v} = x \mathbf{W}_V,
\]

Here, y is the modality indication vector, and x is output tokens from the transformer encoder. \(\mathbf{W}_Q\), \(\mathbf{W}_K\), and \(\mathbf{W}_V\) are learnable weight matrices.
Now, the scaled dot-product cross-attention can be defined as:
\[
\text{attn} = \text{softmax}\left(\frac{\mathbf{q} \mathbf{k}^\text{T}}{\sqrt{d_k}}\right) \mathbf{v}
\]
Where, \(d_k\) is the dimension of the key vectors and the result is scaled by \(\sqrt{d_k}\) to stabilize the gradients during training.

\begin{table}
\caption{Proposed MPViT and ViT embedding dimension and number of heads for different variants }
    \centering
    \begin{tabular}{ c| c| c}
    \hline
        Model & Embedding Dim & Num Heads \\
        \hline
        ViT-tiny & 192 & 3 \\
        ViT-small & 384 & 6 \\
        ViT-base & 768 & 12 \\
        \hline
        MP-ViT-tiny & 192 & 3\\
        MP-ViT-small & 384 & 6\\
        MP-ViT-base & 768 & 12 \\
        \hline
    \end{tabular}
    
    \label{tab:variants_details}
\end{table}

To evaluate the performance of our proposed method quantitatively, we report area under the curve (AUC), sensitivity, and specificity. The AUC indicates the model’s ability to distinguish between positive and negative instances independent of a specific operating point. We compare the performance of MP-ViT with original ViT and traditional CNN architectures, including ResNet \cite{he2016deep} and DenseNet \cite{huang2017densely}. We perform parameter testing to validate the effectiveness of the embedding dimension and the number of attention heads on our proposed MP-ViT and compared it with the original ViT. Different variants of ViT and MP-ViT with embedding dimensions and number of heads are presented in Table \ref{tab:variants_details}.

All the experiments are done on the NVIDIA Tesla V100 SXM2 cluster with 8 GPUS. As the dataset is heavily imbalanced, we use a weighted random sampler to sample the training data. The weighted Adam optimizer is used with learning rate of $1 \times 10^{-4}$ and weight decay of $1 \times 10^{-2}$  with batch size 8. We use cross-entropy loss for the training and each model is trained till 200 epochs and best validation set checkpoint is saved for final testing. All the experiments are implemented using PyTorch(v1.12.1) and the Monai framework \cite{cardoso2022monai}(v.0.8.0).

\section*{Results}



Table \ref{tab:performance} presents the performance on brain hemorrhage classification with multiple anisotropic MR images in different orientations, including axial FLAIR, ADC, Trace, GRE, SWI, T2w and sagittal T1w 3D MR images on a testing dataset with 144 positive and 1352 negative cases.

\begin{figure}[ht!]
\centering
\includegraphics[width=0.65\textwidth]{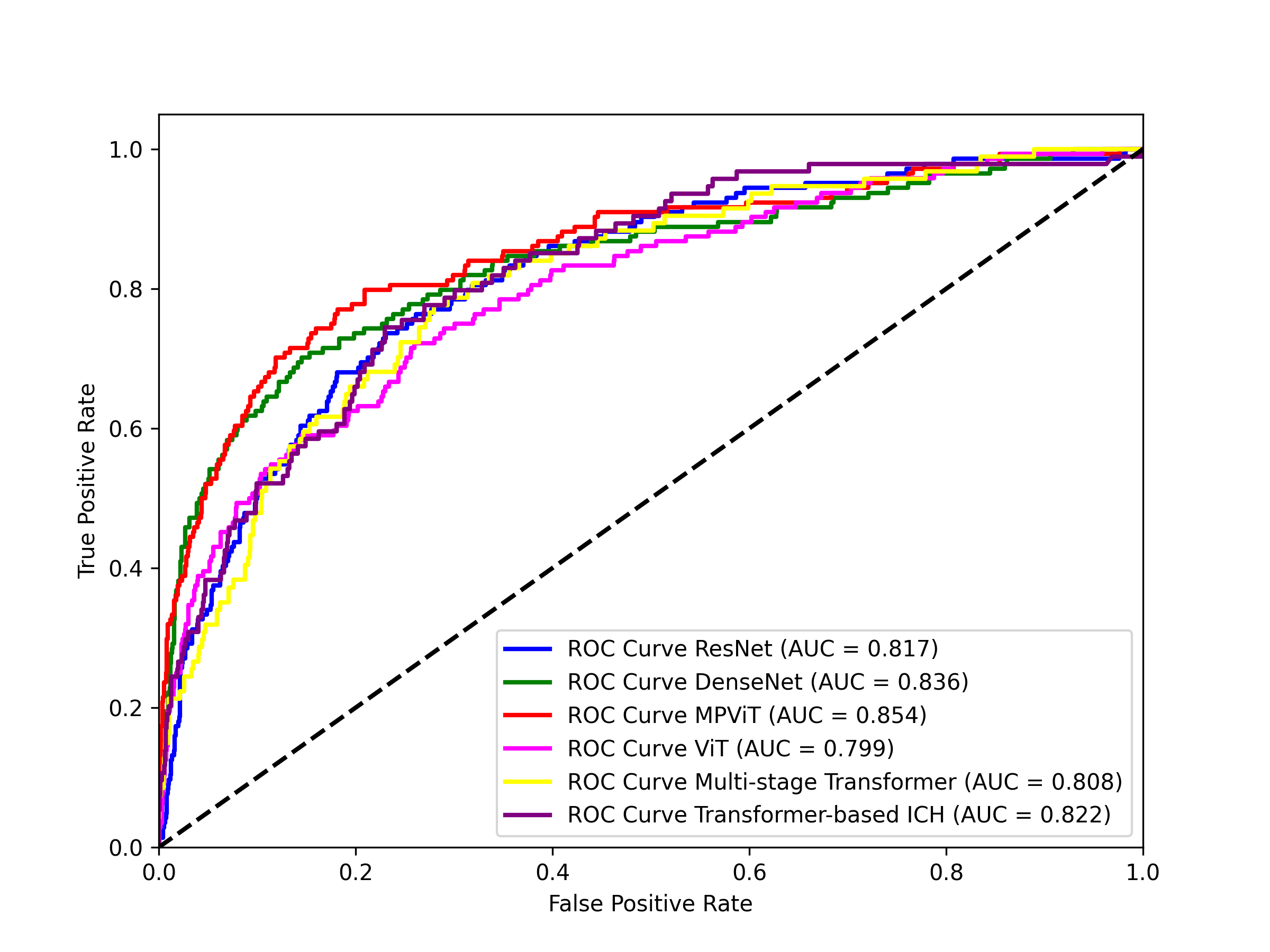}
\caption{Receiver Operating Characteristic (ROC) Curve for different methods for hemorrhage classification} \label{ROC}
\end{figure}

MP-ViT exhibits the highest AUC of 0.854, surpassing ViT (AUC = 0.799) by approximately 5.5\%, ResNet (AUC = 0.817) by approximately 3.7\%, Multi-stage Transformer (AUC=0.808) by 4.6\%, Transformer-based ICH (AUC=0.822) by Adel et al\cite{elzemity2023transformer} by 3.2\%, and DenseNet (AUC = 0.836) by approximately 1.8\%. This suggests that MP-ViT achieves superior overall discriminative performance. The Receiver Operating Characteristic (ROC) curve depicted in Figure \ref{ROC} illustrates the performance of MP-ViT, ViT, ResNet, and DenseNet for hemorrhage classification.

We have three different MP-ViT variants similar to the original ViT with the different configurations presented in Table \ref{tab:performance}. We observe the performance of the base variant with embedding dimension 768 and number of heads 12 achieves the best result.

\begin{table}
\caption{Performance comparison of MP-ViT with ViT and CNN-based models on hemorrhage classification. Here, ResNet101, ResNet152 have 101 and 152 layers respectively. Similarly, DenseNet121, DenseNet169 and DenseNet201 have 121, 169 and 201 layers respectively.}
    \centering
    \begin{tabular}{ c| c| c| c}
    \hline
        Model & AUC & Sensitivity & Specificity\\
        \hline
        ViT-tiny & 0.634 & 0.482 & 0.681\\
        ViT-small & 0.757 & 0.667 & 0.717\\
        ViT-base & 0.799 & 0.632 & 0.783\\
        \hline
       
        ResNet101 & 0.795 & 0.617 & 0.806 \\
        ResNet152 & 0.817 & 0.688 & 0.811 \\
        \hline
        DenseNet121 & 0.819 & 0.513 & 0.963\\
        DenseNet169 & 0.836 & 0.563 & 0.933\\
        DenseNet201 & 0.812 & 0.581 & 0.889\\
        \hline
        MPViT-tiny & 0.775 & 0.926 & 0.495\\
        MP-ViT-small & 0.814 & 0.852 & 0.634\\
        MP-ViT-base & \textbf{0.854} & 0.722 & 0.848\\
        \hline

        Multi-stage Transformer \cite{feng2022multi} & 0.808 & 0.436 & 0.907\\
        Transformer-based ICH \cite{elzemity2023transformer} & 0.822 & 0.479 & 0.908 \\
        \hline

    \end{tabular}
    
    \label{tab:performance}
\end{table}

In the statistical analysis of model comparisons, McNemar’s \cite{mcnemar1947note} test is employed on prediction probabilities of different models. The differences between MP-ViT compared to ViT, ResNet, DenseNet, Multi-stage Transformer and Transformer-based ICH classifier\cite{elzemity2023transformer} are statistically significant (p<0.05) as shown in Table \ref{tab:stats}.

\begin{table}[ht!]
\caption{Statistical comparison of MP-ViT-base with ViT-base and CNN-based models on hemorrhage classification}
    \centering
    \begin{tabular}{ c| c| c }
    \hline
        Model Comparison & p-value & Null Hypothesis Rejected\\
        \hline
        MP-ViT-base Vs ViT-base & $<0.05$ & Yes\\
        \hline
        MP-ViT-base Vs ResNet & $<0.05$ & Yes\\
        \hline
        MP-ViT-base Vs DenseNet & $<0.05$ & Yes \\
        \hline
        MP-ViT-base Vs Multi-stage Transformer\cite{feng2022multi} & $<0.05$ & Yes \\
        \hline
        MP-ViT-base Vs Transformer-based ICH\cite{elzemity2023transformer} & $<0.05$ & Yes \\
        \hline
    \end{tabular}
    
    \label{tab:stats}
\end{table}



    

We perform an ablation study for our MP-ViT with and without the modality indication vector cross-attention module and the performance is reported in Table \ref{tab:vector}. The performance is increased by more than 1\% when we use a modality indication vector which demonstrates model’s robustness in accepting flexible inputs with one or more contrasts absent.

\begin{table}[ht!]
\caption{Performance comparison of MP-ViT with/without modality indication vector. There is a significant difference (p-value<0.05) between these two models.}
    \centering
    \begin{tabular}{c|c}
    \hline
       Model  & AUC\\
    \hline
       MP-ViT-base with Vector  & 0.854\\
       MP-ViT-base without Vector  & 0.842\\
       \hline
    \end{tabular}
    
    \label{tab:vector}
\end{table}

\section*{Discussion}

We propose a multi-plane vision transformer model for hemorrhage classification which can process different plane input data together without changing the orientation. The results of our study demonstrate that the MP-ViT architecture significantly outperforms existing models in brain hemorrhage classification using MR images. MP-ViT achieves an area under the curve (AUC) of 0.854, surpassing the vision transformer (ViT) by 5.5\%, ResNet by 3.7\%, and DenseNet by 1.8\%. MP-ViT also outperformed other hemorrhage classification networks, surpassing the multi-stage Transformer \cite{feng2022multi} by 4.6\% and Adel et al.'s transformer-based ICH classifier \cite{elzemity2023transformer} by 3.2\%. The ablation study further highlights the importance of the modality indication vector. The performance of MP-ViT improved by over 1\% when this vector is included, demonstrating the model’s increased robustness and flexibility in handling cases where one or more contrasts might be missing. This feature allows MP-ViT to maintain high performance even with incomplete data, which is a common issue in real-world clinical settings.

We also observe, all versions of MP-ViT outperform their corresponding ViT counterparts and MP-ViT shows steady improvement as the complexity of the model increases. As MP-ViT has two ViT-based encoders, it has more parameters than the original ViT and hence it is computationally expensive. However, the MP-ViT-small outperforms ViT-base by 1.5\% which has around 48 million less parameters than the later. Similarly, MPViT-Tiny outperforms both ViT-tiny by 14.1\% and ViT-small by 1.8\%. This demonstrates that merely increasing model complexity isn't enough to guarantee better performance; instead, MP-ViT's strength lies in its ability to effectively extract information from multi-plane input images, leading to its superior performance.

In addition, MP-ViT is also well-suited for clinical feasibility. In MR diagnostics, acquiring images in at least two orientations is a common practice, yet most existing methods are not designed to accommodate this requirement. Our approach inherently handles images from different orientations without requiring additional resizing, thereby avoiding unnecessary information loss from downsampling and potential inaccuracies from interpolation. This allows MP-ViT to retain critical diagnostic information, making it particularly effective for multi-orientation MR classification tasks.

It is important to consider the clinical relevance of the results.  While MRI may not be the imaging modality of choice for emergent workup for ICH, it is used in some institutions, particularly in the setting of stroke \cite{fischer2022magnetic}, where the distinction of hemorrhagic vs ischemic stroke is critical, and where hemorrhagic conversion of an ischemic event is also critical to determine.  In such cases, the proposed method can potentially assist the reviewing radiologist in identifying ICH and make a critical difference for the patient.  Beyond the emergency setting, MRI is the imaging modality of choice for many diagnostic workups.  In these non-urgent cases, MRI images may not be reviewed by a radiologist for hours, if not days, after the patient is scanned.  Vernooij et al. have shown as much as a 7\% of adults asymptomatic for ICH have an incidental finding of ICH on MRI scans \cite{vernooij2007incidental}.  In such cases, an AI application which could alert a radiologist of suspected ICH before the patient leaves the scanning facility could potentially have a critical impact on patient outcomes.

It is important to consider a few limitations while interpreting our results. This study is conducted only on an internal dataset for a single task which is hemorrhage classification. Also, it is essential to note the deliberate focus on classification in isolation, excluding the concurrent consideration of segmentation-a critical component often integral for achieving state-of-the-art hemorrhage classification performance \cite{nael2021automated}. This methodological choice is made to isolate the advantages of the MP-ViT classification components over contemporary classification methods which also do not use segmentation. In future, we want to explore methodologies which can be more efficient while maintaining the performance. Additionally, coronal images were not used in this study as they were not available in our dataset, and we aim to explore their inclusion in future work.

\section*{Conclusion}

In this paper, we explore the effectiveness of the Multi-plane vision transformer (MP-ViT) architecture in the context of brain hemorrhage classification using MRI data. MP-ViT outperforms other contemporary models, including vision transformer (ViT), ResNet, and DenseNet, exhibiting the highest area under the curve (AUC) and a balanced combination of sensitivity and specificity. In conclusion, this paper demonstrates a mechanism for vision transformers to incorporate information from images with different orientations and a mechanism for encoding missing contrasts that improves classification performance. These mechanisms should be considered for system designs that must handle the variability of real-world clinical data.





 





\section*{Data availability}
The training, validation, and test datasets used for this study are protected patient information. Some data may be available for research purposes from the corresponding author upon reasonable request.

\bibliography{sample}



\section*{Acknowledgements}

This research project was funded by Siemens Healthineers. We acknowledge the usage of MRI images from the Mount Sinai Hospital.

\section*{Author contributions statement}

E.G., D.C. and A.M. contributed to the conception of the work. B.D., G.Z., and B.M. designed the proposed architecture. B.D. implemented multi-plane Vision Transformer for experiments and conducted the experiments. T.J.R. performed data interpretation and infarct segmentation. G.Z., E.G., and A.M. analyzed the results. B.D. and G.Z. wrote the manuscript. E.G., A.M., and D.C. secured funding. All authors reviewed the manuscript. 

\section*{Funding}
Open Access funding enabled and organized by Projekt DEAL.

\section*{Competing interests}
B.D., G.Z., E.G., B.M., and D.C. are employed by Siemens Healthineers. T.J.R. is a paid consultant for Siemens Healthineers. All other authors declare no competing interest.

\section*{Additional information}
Correspondence and requests for materials should be addressed to B.D.

\section*{Code availability}
The code base for the deep-learning framework makes use of proprietary components and we are unable to publicly
release the code base. However, all experiments and implementation details are described in sufficient
detail in the Methods to enable independent replication with non-proprietary libraries.




\end{document}